\documentclass{article}
\usepackage{latexsym,amssymb, amsfonts, amscd}

\newtheorem{lemma}{Lemma}
\newtheorem{remark}{Remark}

\newtheorem{theorem}{Theorem}

\begin{document}
\title{Analysis of the Faddeev model}
\author{Dave Auckly \thanks{{The first author was partially supported by
NSF grant DMS-0204651.}}
\and Lev Kapitanski
\thanks{{  
The second author was partially supported by NSF grant
DMS-0200670.} }%
}                     
\date{
$$
\begin{array}{cc}
\hbox{Department of Mathematics} & \hbox{Department of Mathematics}\\ 
\hbox{Kansas State University} & \hbox{University of Miami}\\ 
\hbox{Manhattan, KS 66506, USA } & \hbox{Coral Gables, FL 33124, USA}
\end{array}
$$
}
%
%
%
%
\maketitle

\begin{abstract} 
In this paper we consider a generalization of the Faddeev  model for the maps from 
a closed three-manifold into the two-sphere. We give a novel representation of 
smooth $\,S^2$-valued maps based on flat connections. This representation 
allows us to obtain an analytic description of the homotopy classes of 
$\,S^2$-valued maps that generalizes to Sobolev maps. 
It also leads to a new proof an old theorem of Pontrjagin. 
For the generalized Faddeev 
model,
we prove the existence of minimizers  in every homotopy class. 
\end{abstract}
%
\section{Introduction}
\label{intro}
In 1975 L. D. Faddeev introduced
 an interesting nonlinear sigma-model motivated by the Hopf invariant
and the Skyrme model, \cite{Faddeev}.
The Hopf invariant is an integer associated to any continuous map from
the $3$-sphere to the $2$-sphere. Roughly, it counts the linking
number of the inverse images of two generic points on $\,S^2$.
Homotopy classes of such maps are
completely classified by the Hopf invariant, \cite{Hopf,Bott-Tu}.
Faddeev's model is now often called the Faddeev - Hopf model. 
It is also referred to as the Faddeev - Skyrme model.
The fields in this model are maps $\,{\bf n}\,$
from $\,\mathbb R^3\,$ to $\,S^2\,$
asymptotically constant at infinity. The energy functional is
\begin{equation}\label{1}
E({\bf n})\,=\,\int_{\mathbb R^3} |d{\bf n}|^2\,
+\,|d{\bf n}\wedge d{\bf n}|^2\,d^3x\,.
\end{equation}
The Hopf invariant can be evaluated analytically
as follows, \cite{Bott-Tu},
\begin{equation}\label{hopf}
Q({\bf n})\,=\,\int_{\mathbb R^3} \alpha\wedge d\alpha\,,
\end{equation}
where $\,\alpha\,$ is the unique $\,\delta$-closed $1$-form 
vanishing at infinity 
such that $\,d\alpha\,$ equals the pull-back, $\,{\bf n}^*\omega_{S^2}$,
of the volume $2$-form  on $\,S^2$. In coordinates,
$\,\omega_{S^2}\,=\,\frac{1}{4\pi}\left(
n^1\,dn^2\wedge dn^3\,+\,n^2\,dn^3\wedge dn^1\,+n^3\,dn^1\wedge dn^2\right)\,$. 
In the homotopy class corresponding to
$\,Q({\bf n})\in \mathbb Z$,
the energy is bounded from below (see \cite{VK}):
\[
E({\bf n})\,\ge\,const\cdot |Q({\bf n})|^{\frac 34}\;.
\]
This estimate leads one to believe that each sector (homotopy class)
should have a ground state -- a minimizer of $\,E({\bf n})$.
Faddeev expected the minimizers with $\,Q({\bf n})\neq 0\,$
to have interesting knot-like structures, \cite{Faddeev},
and recently several numerical investigations have
provided some support to this conjecture, see \cite{Faddeev2,Faddeev-Niemi}
and the references therein. Mathematically, one problem 
with a domain of  $\,\mathbb R^3\,$ is that it is not compact; 
intuitively, one may imagine a minimizing sequence with 
a concentrated lump sliding to infinity. When the domain is a closed 
three-manifold, this particular difficulty is avoided. 
However, all of the other interesting physical and mathematical features 
remain. If one only wished to consider minimizers on $\,S^3$, 
the Hopf invariant gives the homotopy classification and has 
an analytic expression analogous to (\ref{hopf}).

When $\,\mathbb R^3\,$ or $\,S^3\,$ is replaced by a general Riemannian 
three-manifold, $\,M^3$, the homotopy classification of
maps to $\,S^2\,$ is more complicated. The classification result
is due to Pontrjagin, \cite{Pontrjagin}. Pontrjagin, in fact,
classifies the maps from general three-complexes to the $2$-sphere.
For the special case of three-manifolds, we repeat his result in the
proposition below. Compared to the $\,S^3\,$ case, two new features
arise. First of all, there is a new invariant given by the induced
map on second cohomology. Second, the Hopf invariant generalizes into a
secondary invariant that sometimes takes values in a finite cyclic group.

\begin{theorem}[Pontrjagin]\label{thm-1}
 Let $\,M\,$ be a closed, connected, oriented
three-manifold. To any continuous map $\,\phi\,$ from $\,M\,$
to $\,S^2\,$ one associates a cohomology class $\,\phi^*\mu_{S^2}\in H^2(M;\mathbb Z)$,
where $\,\mu_{S^2}\,$ is a generator of $\,H^2(S^2;\mathbb Z)$.
Every class may be obtained from some map, and two maps with
different classes lie in different homotopy classes.
The homotopy classes of maps with a fixed class $\,\alpha\in H^2(M;\mathbb Z)\,$
are in bijective correspondence with
$\,H^3(M; \mathbb Z)/(2\alpha\cup H^1(M;\mathbb Z))$.
\end{theorem}

The known proofs of Pontrjagin's proposition provides a pretty picture 
of the homotopy classification. These proofs are geometric and 
combinatorial in nature, and cannot be easily used in our 
minimization problem. To circumvent this difficulty, we give a novel 
description of smooth maps from closed three-manifolds into the 
$2$-sphere. Namely, an $\,S^2$-valued map will be represented  
by a flat connection and a smooth reference map. Here we rely 
on our earlier work \cite{AK} and on the concurrent research 
in \cite{AS}. 
This description allows us to obtain an analytically 
friendly picture of the homotopy classification and a new proof of 
Pontrjagin's result. At the same time, this presentation fits in well 
with Faddeev's functional, which we rewrite using the connection and reference 
map. In \cite{AS} the developing map associated with the connection 
is used to  compute the fundamental group and rational cohomology 
of the configuration space.   

The first part of Section 2 relates maps into $\,S^2$ 
to maps into $\,S^3$ which is the basis of our proof of the
Pontrjagin  theorem. We then introduce flat connections to 
encode $\,S^3$-valued maps. To encode an $\,S^3$-valued map by a flat 
connection, a framing is required. Without the framing, the map 
is determined up to an orientation preserving isometry of $\,S^2$. 
Since the Faddeev functional is invariant under isometries, 
we ignore framings. Our description of orientation preserving 
isometry classes of smooth $\,S^2$-valued maps in terms of special 
flat connections is given in Theorem 2. 
In Section 3 we rewrite the Faddeev energy functional in terms of flat 
connections. At this point we turn from smooth connections to connections 
with finite energy. Our main result, Theorem 3, is that the analytical 
conditions fixing the homotopy class are well-defined for finite energy 
connections. In addition, we prove the existence of a minimizer 
of the energy functional in every class. 
When the primary obstruction vanishes, there is an alternate approach 
to minimization that may be interesting in its own right. This is 
discussed in Section 4.


\section{$S^2$-valued maps and homotopy classification}\label{sec:2}

Let $\,M\,$ be a closed, orientable 3-manifold and $\,\mu_{S^2}\,$ 
be a generator of $\,H^2(S^2,\mathbb Z)$. The image of $\,\mu_{S^2}\,$ 
in the de Rham cohomology is the equivalence class of the form 
$\,\omega_{S^2}\,$ given previously.
For smooth (or just continuous) maps  $\,\psi:\,M\to S^2$, it is well 
known that $\,\psi^*\mu_{S^2}\,$ is a homotopy invariant. Every 
class $\,\alpha\in H^2(M,\mathbb Z)\,$ arises from some map $\,\psi$. 
Here is the standard construction:
Let $\,\gamma\,$ be a $1$-cycle in $\,M\,$ dual 
to $\,\alpha$. Since $\,M\,$ is orientable, the normal bundle to 
$\,\gamma\,$ is trivial. Using a trivialization of the normal bundle, 
each fiber may be identified with $\,\mathbb R^2\,$ and mapped via stereographic 
projection onto the $2$-sphere. Finally, map the complement of the normal bundle 
to the North pole.  Now, there are many trivializations of 
the normal bundle. Any two trivializations are related by some number 
of twists (full rotations of the fiber when moving along $\,\gamma$). 
The number of twists is the secondary invariant described in the second  
part of Proposition \ref{thm-1}. To describe the secondary 
invariant analytically, we will need a few constructions. 
We start with notation. 

\medskip

\noindent{\bf Notation.} 
\begin{description}
\item $\bullet$\ \ $\,\hbox{Sp}_1\,$ is the group of unit quaternions, 
$\,q^*\,$ denotes the quaternionic conjugation of $\,q$. The Lie algebra of 
$\,\hbox{Sp}_1\,$ is the purely imaginary quaternions, denoted $\,\mathfrak{sp}_1$. 
\item $\bullet$\ \ $\,S^1\,$ is the group of unit complex numbers 
regarded as a subgroup of $\,\hbox{Sp}_1$.
\item $\bullet$\ \ $\,S^2\,$ will be identified with the unit sphere in the 
purely imaginary quaternions.
\item $\bullet$\ \ The usual dot-product may be expressed using quaternionic 
multiplication as $\,\langle p,\,q\rangle\,=\,\frac{1}{2}(p^* q\,+\,q^*\,p)$.
\item $\bullet$\ \ $C^\infty(X,Y)\,$ denotes the space of smooth maps 
from $X$ to $Y$.
\item $\bullet$\ \ $\,W^{s,p}\,$ denotes the usual Sobolev space of functions 
with $s$ derivatives in $\,L^p$; $\,W^{s,p}(M, \hbox{Sp}_1)\,$ denotes the subset 
of quaternion-valued $\,W^{s,p}\,$ functions on $\,M\,$ which take values in 
$\,\hbox{Sp}_1\,$ almost everywhere; $\,W^{s,p}(M, S^2)\,$ is defined analogously.

\end{description}

Given any map $\,\varphi:\,M\to S^2\,$ and any map $\,u:\,M\to \hbox{Sp}_1$, 
we construct a new map $\,\psi:\,M\to S^2\,$ by $\,\psi(x)=u(x)\varphi(x) u(x)^*$. 
We will show that $\,\psi\,$ has the same associated cohomology class as 
$\,\varphi$. Conversely, we will see that any map $\,\psi\,$ with the same 
associated cohomology class may be represented in this way. 

To prove these facts we will need several maps (compare with the discussion 
in \cite{AS}). The most important map 
is $\,\mathfrak q:\,S^2\times S^1\to \hbox{Sp}_1\,$ defined by 
$\,\mathfrak q(z,\lambda)\,=\,q\,\lambda\,q^*$, where $\,z\,=\,q\,i\,q^*$. 
This bizarre looking map will later encode the gauge freedom when we describe 
$\,S^2$-valued maps via $\,S^3$-valued maps. 
It will be important later that $\,\mathfrak q\,$ has degree $2$ (with 
standard orientations). One can check that $\,i\,$ is a regular value 
with inverse image $\,\{\pm (i, i)\}$, and $\,\mathfrak q\,$ is orientation 
preserving at each point. 
The second map is given by 
$\,f:\,S^2\times \hbox{Sp}_1\to S^2\times S^2$, where $\,f(z,q)\,=\,(z,q\,z\,q^*)$.
Define a free right $\,S^1$-action 
$\,\rho:\,S^2\times \hbox{Sp}_1\times S^1\to S^2\times\hbox{Sp}_1\,$ 
by $\,\rho(z,p,\lambda)\,=\,(z,p\,\mathfrak q(z,\lambda))$.  The quotient 
map associated with $\,\rho\,$ is $\,f$. Thus $\,f\,$ is the projection 
of a principal fiber bundle by standard facts from the theory of group actions on manifolds, 
and therefore, $\,f\,$ is a fibration by the covering homotopy theorem, \cite{Bredon,Spanier}. For comparison, recall that the Hopf fibration 
$\,h:\,\hbox{Sp}_1\to S^2\,$ is 
given by the map
$\,h(q)\,=\,q\,i\,q^*$.

Consider two maps, $\,\varphi\,$ and $\,\psi$, going from $\,M\,$ to $\,S^2$. 
Define $\,Q_{\varphi,\psi}\,=\,\{(x,q)\in M\times \hbox{Sp}_1\,|\,
\psi(x)\,=\,q\,\varphi(x)\,q^*\,\}$. This bundle is the pull-back of the 
fibration $\,f\,$ under the map $\,(\varphi,\psi):\,M\to S^2\times S^2$. 
\begin{lemma} \label{lemma1}
There exists a smooth map $\,u:\,M\to \hbox{Sp}_1\,$ such that 
 $\,\psi\,$ and $\,\varphi\,$ are related by $\,\psi\,=\,u\,\varphi\,u^*\,$ 
if and only if  
$\,\psi^*\mu_{S^2}\,=\,\varphi^*\mu_{S^2}$. 
\end{lemma}

\noindent{\bf Proof.} Denote $\,P\,=\,S^2\times\hbox{Sp}_1\,$ considered as a 
principal bundle over $\,S^2\times S^2\,$ with bundle map $\,f$.
Let $\,f_1,\,f_2:\,P\,\to\,S^2\,$  be the first and second components of $\,f$,
i.e., $\,f_1(z,q)\,=\,z$ and 
$\,f_2(z,q)\,=\,q\,z\,q^*$.  Notice, that $\,f_2^*\mu_{S^2}\,=\,f_1^*\mu_{S^2}\,$ 
because $\,f_2^*\mu_{S^2}[S^2\times\{1\}]\,=\,f_1^*\mu_{S^2}[S^2\times\{1\}]\,$ 
and $\,[S^2\times\{1\}]\,$ generates $\,H_2(S^2\times\hbox{Sp}_1;\mathbb {Z})$.
If $\,\psi\,=\,u\,\varphi\,u^*$, then 
$\,\psi^* \mu_{S^2}\,=\,(\varphi, u)^*\,f_2^*\,\mu_{S^2}\,
=\,(\varphi, u)^*\,f_1^*\,\mu_{S^2}\,=\,\varphi^* \mu_{S^2}$. 

In the opposite direction, assume that $\,\psi^*\mu_{S^2}\,=\,\varphi^*\mu_{S^2}$. 
We will prove that the bundle $\,Q_{\varphi,\psi}\,$ is then trivial. 
Hence, it admits a 
section $\,\sigma: M\to Q_{\varphi,\psi}$. The composition 
of $\,\sigma\,$ with the projection on the second component of
$\,Q_{\varphi,\psi}\,$ 
is the desired $\,u$. To see that $\,Q_{\varphi,\psi}\,$ is trivial, we compute the first Chern class of the associated line bundle.
We have 
\begin{equation}
\begin{array}{rl} 
c_1(Q_{\varphi,\psi})\, = &\,(\varphi,\psi)^*c_1(P)\,\\
= &\,(\varphi,\psi)^*(\hbox{pr}_1^*\,\mu_{S^2}\,-\,\hbox{pr}_2^*\,\mu_{S^2})\\ 
= &\,\varphi^*\mu_{S^2}\,-\,\psi^*\mu_{S^2}\,=\,0.
\end{array}
\end{equation}
  In the above computation, $\,\hbox{pr}_k\,$ 
is the projection of $\,S^2\times S^2\,$ onto the $k$-th factor. 
We have $\,c_1(P)[\{\mathbf i\}\times S^2]\,=\,-1$, since 
$\,f\,$ restricted to $\,\{\mathbf i\}\times S^2\,$ is the Hopf fibration. 
In addition, for the diagonal map $\,\Delta:\,S^2\to S^2\times S^2\,$ 
we have $\,c_1(P)[\Delta(S^2)]\,=\,0$, since $\,\Delta^*P\,$ admits the section 
$\,\sigma(z)\,=\,(z,1)$. Recalling that $\,[\Delta(S^2)]=[\{\mathbf i\}\times S^2] + [S^2\times\{\mathbf i\}]$, we conclude that 
$\,c_1(P)\,=\,\hbox{pr}_1^*\,\mu_{S^2}\,-\,\hbox{pr}_2^*\,\mu_{S^2}$, as 
used above. \hfill$\square$

We can now complete our proof of Pontrjagin's theorem. 
Fix a reference map $\,\varphi$. Let $\,\mathfrak D(M, S^2)\,$ be the set of smooth
maps from 
$\,M\,$ to $\,S^2\,$ with
the same associated cohomology class as $\,\varphi$. 
Using the covering homotopy property of the fibration $\,f$, 
we obtain a fibration $\,C^\infty(M,\hbox{Sp}_1)\to \mathfrak D(M, S^2)\,$ 
with homotopy fiber $\,C^\infty(M, S^1)$. The fiber is included by 
$\,\lambda\mapsto \mathfrak q(\varphi,\lambda)$. The fibration is given by 
$\,u\mapsto u\,\varphi\,u^*$. This fibration induces a short exact sequence 
at the level of path components:
$$
\pi_0(C^\infty(M,S^1))\to \pi_0(C^\infty(M,\hbox{Sp}_1))
\to \pi_0(\mathfrak D(M,S^2))\,.
$$
It is well known that $\,\pi_0(C^\infty(M,S^1))\,$ is isomorphic to 
$\,H^1(M;\mathbb Z)\,$ by $\,\lambda\mapsto \lambda^*\mu_{S^1}$, 
and $\,\pi_0(C^\infty(M,\hbox{Sp}_1))\,$ is isomorphic to 
$\,H^3(M;\mathbb Z)\,$ by $\,u\mapsto u^*\mu_{\hbox{Sp}_1}$. 
Now, 
\begin{equation}\label{q*}
\mathfrak q(\varphi, \lambda)^*\mu_{\hbox{Sp}_1}\,=\,
(\varphi, \lambda)^*\mathfrak q^*\mu_{\hbox{Sp}_1}\,=\,
(\varphi, \lambda)^*(2\mu_{S^2}\cup \mu_{S^1})\,=\,
2 \varphi^*\mu_{S^2}\cup \lambda^*\mu_{S^1}).
\end{equation}
Here we used the fact that $\,\mathfrak q\,$ has degree $2$. 
From this computation and the exact sequence, it follows that 
$\,\pi_0(\mathfrak D(M,S^2))\,\cong\,H^3(M;\mathbb Z)/
\,(2\varphi^*\mu_{S^2}\cup H^1(M;\mathbb Z)$. 
This completes our proof of Pontrjagin's theoerem. 

\medskip

Fix a reference map $\,\varphi:\,M\to S^2$. 
We have just shown that any map $\,\psi\,$ with the same associated 
cohomology class $\,\varphi^*\mu_{S^2}\,$ may be represented in the form 
$\,\psi\,=\,u\,\varphi\,u^*\,$ for some $\,u:\,M\to \hbox{Sp}_1$. 
If $\,\psi\,$ is homotopic to $\,\varphi$, then, in addition, 
$\,u\,$ may be chosen homotopic to a constant map. This follows by a simple 
application of the covering homotopy property of the fibration $\,f$. 
In fact, there are many such maps $\,u$.  
For  any  map $\,\lambda:\,M\to S^1$, the map 
$\,u\,\mathfrak q(\varphi,\lambda)\,$ may also be used to represent $\,\psi$. 
However, $\,u\,\mathfrak q(\varphi,\lambda)\,$ will not necessarily be 
null-homotopic. Varying the map $\,\lambda\,$ one obtains all 
$\,\hbox{Sp}_1$-valued maps representing $\,\psi$. 
Using equation (\ref{q*}) we see that 
\begin{equation}\label{2d}
\hbox{deg}\ u\,{\frak q}(\varphi,\lambda)\,=\,
\left(2\,\varphi^*\mu_{S^2}\cup \lambda^*\mu_{S^1}\right)\,[M]\,+\,\hbox{deg}\,u\,,
\end{equation}
 The term 
$\,\left(2\,\varphi^*\mu_{S^2}\cup \lambda^*\mu_{S^1}\right)\,[M]\,$ 
in (\ref{2d}) 
is an even integer, because $\,\mu_{S^2}\,$ and $\,\mu_{S^1}\,$ are integral classes. 
The map $\,\eta \mapsto (\varphi^*\mu_{S^2}\cup \eta)[M]\,$ from 
$\,H^1(M;\mathbb Z)\,$ to $\,\mathbb Z\,$ is a group homomorphism, 
and therefore has image $\,m\,\mathbb Z\,$ for some $\,m\,$ depending on 
the class $\,\varphi^*\mu_S^2$.
 Since the degree of a null-homotopic 
map is zero, we conclude, that the degree of any map $\,u\,$ corresponding to 
a map $\,\psi\,$ homotopic to $\,\varphi\,$ lies in $\,2\,m\,\mathbb Z$.

\begin{remark} All homotopy classes of maps $\,\psi:\,M\to S^2\,$ with the same 
second cohomology class $\,\psi^*\mu_{S^2} = \varphi^*\mu_{S^2}\,$ are obtained 
in the form $\,\psi\,=\,u\,\varphi\,u^*$. The maps $\,u_1\,\varphi\,u_1^*\,$ 
and $\,u_2\,\varphi\,u_2^*\,$ are homotopic if and only if 
$\,\hbox{deg}\,u_1\,\equiv \hbox{deg}\,u_2\,(\hbox{mod}\; 2\,m)$.
\end{remark}

Every map $\,u:\,M\to \hbox{Sp}_1\,$ 
is the developing map of a flat connection $\,a\,=\,u^{-1}du$. 
Connections arising from such $\,u$'s have trivial holonomy. Conversely, 
given any flat connection, $\,a$, with trivial holonomy, one can find a 
map $\,u:\,M\to \hbox{Sp}_1\,$ so that $\,a\,=\,u^{-1}du$. Such $\,u\,$ is 
unique up to left multiplication by a constant in $\,\hbox{Sp}_1$.
Left multiplication of $\,u\,$ by a constant will change $\,\psi\,$ by an 
orientation preserving isometry in $\,S^2$. 
In general, the degree of the map $\,u\,$ is given by 
$\,-\frac{1}{12\pi^2}\,\int_M \hbox{Re} (a\wedge a\wedge a)$. Notice that $\, da+a\wedge a\, =\, 0$\, implies that this integral 
is exactly the Chern-Simons invariant, 
$$ \hbox{cs}(a)\, = \, \frac{1}{4\pi^2}\int_M \hbox{Re}(a\wedge da+\frac23 a\wedge a \wedge a),
$$
of the flat
connection $\,a$. 
Recall from the previous paragraph 
that $\,u\,\varphi\,u^*\,$ is homotopic to $\,\varphi\,$ if and only if  
$\hbox{deg}\,u\,=\,-\frac{1}{12\pi^2}\int_M \hbox{Re} (a\wedge a\wedge a)\,\in\,2m\mathbb Z$.

For any $\,\psi = u\,\varphi\,u^*\,$ homotopic to $\,\varphi$, we consider the Hodge
decomposition  of the  
$\,\varphi$-component, $\,\langle a,\,\varphi\rangle$, of the associated connection
$\,a = u^{-1}\,du$. 
Let $\,{\cal H}\langle a,\,\varphi\rangle\,$ be  
the harmonic component of $\,\langle a,\,\varphi\rangle$. 
The space of harmonic $1$-forms on $\,M\,$ is identified with 
$\,H^1(M;\mathbb R)$. Let $\,\{\eta_1,\dots, \eta_k\}\,$ be an integral basis 
for $\,H^1(M;\mathbb R)\,$ and write 
$$
{\cal H}\langle a,\,\varphi\rangle\,=\,h_1\,\eta_1+\dots +
h_b\,\eta_b\,.
$$ 
Recall, that every class 
$\,\eta\in H^1(M;\mathbb Z)\,$ is a pull-back of $\,\mu_{S^1}\,$ 
under a smooth map $\,\lambda:\,M\to S^1$. Given such $\,\lambda$, 
let $\,a^{\lambda}\,=\,\mathfrak q^{-1}\,a\,\mathfrak q\,
+\,\mathfrak q^{-1}\,d\mathfrak q\,$ be the flat connection corresponding 
to $\,u\,{\mathfrak q}(\varphi,\lambda)$. 
Notice, that 
$\,
\langle a^{\lambda},\,\varphi\rangle\,-\,
\langle a,\,\varphi\rangle\,$ is a de Rham representative of the 
 image of $\,\lambda^*\mu_{S^1}\,$ in $\,H^1(M;\mathbb R)$.
We conclude, that by an appropriate choice of gauge, $\,\lambda$, we can make 
each coefficient 
$\,h_k\,$ 
assume values in the interval
$\,[0,1)$. After this is achieved, turn to the $\,d$-component of 
$\langle a,\,\varphi\rangle$. Note that 
$\,u\,\mathfrak q (\varphi, e^{i\theta})\,$ will also represent $\,\psi$ 
for any smooth $\,\theta:\,M\to\mathbb R$. The flat connection corresponding to 
$\,u\,\mathfrak q(\varphi, e^{i\theta})\,$ is 
$\,a^{\exp(i\theta)}\,=\,\mathfrak q^{-1}\,a\,\mathfrak q\,
+\,\mathfrak q^{-1}\,d\mathfrak q$. Now, 
$\,
\langle a^{\exp(i\theta)},\,\varphi\rangle\,-\,
\langle a,\,\varphi\rangle\,=\,d\theta$.
Thus, by an appropriate choice of gauge, we may further 
assume that the
$\,d$-component of 
$\,\langle a,\,\varphi\rangle\,$ is zero. 
We summarize all of these observations in the following theorem.

\begin{theorem}\label{Thm2} 
Any orientation preserving 
$\,S^2$-isometry class of a smooth map
from
$\,M\,$  to $\,S^2\,$ homotopic to $\,\varphi\,$ is uniquely represented 
by a smooth flat connection $\,a$, which has trivial holonomy and satisfies the
conditions 
\begin{enumerate}
\item $\hbox{cs}(a)\,=\,-\frac{1}{12\pi^2}\,\int_M \hbox{Re} (a\wedge a\wedge a)\,\in\,2 m\mathbb Z=\{2(\varphi^*\mu_{S^2}\cup \eta)[M]\,\vert\,\eta\in H^1(M;\mathbb Z)\}$.
\item ${\cal H}\langle a,\,\varphi\rangle\,=\,h_1\,\eta_1+\dots +
h_b\,\eta_b\,$ with $\,h_1,\dots, h_b\in [0,1)$.
\item $\delta\,\langle a,\,\varphi\rangle\,=\,0$\ (here and below 
$\,\delta\,$ is the
adjoint  of the exterior derivative, $d$).
\end{enumerate}
\end{theorem}


\section{Minimizers of the Faddeev functional, (I)}
\label{sec:3} 

The Faddeev energy of a map $\,\psi:\,M\to S^2$, 
$\,\psi(x)\,=\,\psi^1(x)\,{\bf i} + \psi^2(x)\,{\bf j} + \psi^3(x)\,{\bf k}$, is
given by 
$$
E(\psi)\,=\,\int_M |d\psi|^2\,+\,|d\psi\wedge d\psi|^2\,d\hbox{vol}\,,
$$
where 
$$
|d\psi|^2\,=\,|d\psi^1|^2\,+\,|d\psi^2|^2\,+\,|d\psi^2|^2\,,
$$
and 
$$
\begin{array}{rl}
|d\psi\wedge d\psi|^2\,= &\,|d\psi^1\wedge d\psi^2|^2\,+\,|d\psi^2\wedge d\psi^3|^2\,
+\,|d\psi^3\wedge d\psi^1|^2\\ 
= &\,\frac{1}{4}\,|[d\psi, d\psi]|^2.
\end{array}
$$
We begin this section by rewriting $\,E(\psi)\,$ using the representation 
of $\,S^2$-valued maps in Theorem \ref{Thm2}. Fix a smooth reference map 
$\,\varphi :\,M\to S^2$. If $\,\psi\,$ is smooth and homotopic to $\,\varphi$, 
it can be represented as $\,\psi\,=\,u\,\varphi\,u^*\,$ with 
$\,u:\,M\to\hbox{Sp}_1$. Substitute this expression into the energy functional, 
use $\,\hbox{Ad}$-invariance of the norm and the Lie bracket, and the notation 
$\,a = u^*\,du\,$ to
obtain
\begin{equation}\label{E}
E(\psi)\,=\,E_{\varphi}[a]\,:=\,\int_M |D_a \varphi|^2\,+\,
|D_a \varphi\wedge D_a \varphi|^2\,,
\end{equation}
where $\,D_a \varphi\,=\,d\varphi\, +\,[a,\varphi]$.  
There are several advantages in using $\,E_\varphi[a]\,$ as the primary 
expression for the energy functional, the two main advantages are: the conditions 
fixing the homotopy class can be expressed analytically in terms of $\,a$; 
the primary field, $\,a$, takes values in a linear space. 
The natural space for our minimization problem is the space 
$\,\frak A_\varphi\,$ of finite energy flat connections $\,a\,$ satisfying 
the conditions of Theorem \ref{Thm2}. More precisely, we assume that 
$\,a\in L^2(M, \mathfrak{sp}_1)$, that $\,da+a\wedge a=0\,$ in the sense of
distributions, and that $\,a\,$ has trivial holonomy, $\,\rho_a\,=\,1$ 
(see \cite{AK}, Section 3,  Lemma 4). In addition, we assume that 
$\,E_\varphi[a]<\infty\,$ and 
$\,\hbox{cs}(a)\,=\,-\frac{1}{12\pi^2}\int_M \hbox{Re} (a\wedge a\wedge a)\,\in 2 m\mathbb Z$.  
Also, we require that $\,{\cal H}\langle a,\,\varphi\rangle
\,=\,h_1\,\eta_1+\dots +
h_b\,\eta_b\,$ with $\, h_k\in [0,1]\,$ 
and $\,\delta\,\langle a,\,\varphi\rangle\,=\,0$. Denote this class by 
$\,\frak A_\varphi $. Note, that we now allow $\,h_k=1$. By doing so, we lose 
the unique representation of the orientation preserving isometry class 
of $\,\psi$, but this is more convenient for taking limits, 
and we can always return to the case $\,h_k\in [0,1)\,$ by a smooth gauge
transformation.

\begin{theorem}\label{theorem3} The functional $\,E_\varphi[a]\,$ has 
a minimum in the class $\,\frak A_\varphi $.
\end{theorem}

\medskip

\noindent{\bf Proof.\/} Let $\,a_n\,$ be a minimizing sequence in 
$\,\frak A_\varphi $. Our first step is to show that $\,a_n\,$ are uniformly 
bounded
in $\,L^2$. For each of the forms $\,a_n\,$ we use the orthogonal 
decomposition
\[
a_n\,=\,\langle
a_n,\,\varphi\rangle\,\varphi\,
+\,\frac12\,\varphi\,[a_n,\,\varphi]\,.
\]
Accordingly, the curvature condition, $\,da_n + a_n\wedge a_n = 0$, decomposes into
\begin{equation}\label{eq1}
d\;\langle a_n,\,\varphi\rangle\,=\,\frac14\,\left[\varphi\,d\varphi, \, [a_n,\varphi]\right]\,
+\,\frac14\,\varphi\,[a_n,\varphi]\wedge [a_n,\varphi]\,,
\end{equation}
\begin{equation}\label{eq2}
d\;\left(\frac12\,\varphi[a_n,\varphi]\right)\,= \,\langle a_n,\,\varphi\rangle\wedge
D_{a_n}\varphi\, +\,\frac14\,d\varphi\wedge [a_n,\varphi]\,+\,\frac14\,[a_n,\varphi]\wedge d\varphi\, .
\end{equation}
Recall that $\,\varphi\,$ is a smooth $\,S^2$-valued function. 
The terms $\,[a_n,\,\varphi]\,$ and $\,[a_n,\,\varphi]\wedge [a_n,\,\varphi]\,$ are
uniformly bounded in $\,L^2$:
$$
\| [a_n,\,\varphi]\|_{L^2}\,\le\,\| D_{a_n}\varphi\|_{L^2}\,+\,\|d\varphi\|_{L^2}\;,
$$
$$
\|[a_n,\,\varphi]\wedge [a_n,\,\varphi]\|_{L^2}\,\le\,
\|D_{a_n}\varphi\wedge D_{a_n}\varphi\|_{L^2}\, +\,
2\,\|d\varphi\|_{L^\infty}\,\| [a_n,\,\varphi]\|_{L^2}\,+\,\|d\varphi\wedge d\varphi\|_{L^2}\,.
$$
The harmonic part of $\,\langle a_n,\,\varphi\rangle\,$ is uniformly 
bounded in
$\,L^\infty\,$  by the assumptions of our class $\,\frak A_\varphi$. 
From equation (\ref{eq1}) and $\,\delta\,\langle a_n,\,\varphi\rangle\,=\,0\,$ 
we obtain an elliptic estimate
\begin{equation}\label{elliptic}
\|\langle a_n,\,\varphi\rangle\|_{W^{1,2}}\,\le\,
C\left(\|\,d\,\langle a_n,\,\varphi\rangle\|_{L^2}\,+\,
\|{\cal H}\langle a_n,\,\varphi\rangle\|_{L^2}\right)\,.
\end{equation} 
Thus, the sequence $\,a_n\,$ is uniformly bounded in $\,L^2$. Choose a 
subsequence, called also $\,a_n$, 
that converges weakly in $\,L^2$, and let the limit be
$\,a$.  Note that $\,\langle a_n,\,\varphi\rangle\,\varphi\rightharpoonup 
\langle a,\,\varphi\rangle\,\varphi\,$ and 
$\,\varphi\,[a_n,\varphi]\rightharpoonup \varphi\,[a,\varphi]\,$ in $\,L^2$. 
In fact, 
$\,\langle a_n,\,\varphi\rangle\,$ converges to 
$\,\langle a,\,\varphi\rangle\,$ strongly in $\,L^p\,$ for $\,2\le p<6\,$
by the compactness of the embedding $\,W^{1,2}\subset L^p$. 
 Consider the wedge products
$\,\varphi\,[a_n,\varphi]\wedge \varphi\,[a_n,\varphi]$.  By sparsing the sequence
$\,a_n$, we will assume that 
$\,\varphi\,[a_n,\varphi]\wedge \varphi\,[a_n,\varphi]\,$ converges weakly in
$\,L^2\,$ to some $2$-form $\,\xi$. Let us show that 
$\,\xi\,=\,\varphi\,[a,\varphi]\wedge \varphi\,[a,\varphi]$. 
We need the following 
version of the {\it div - curl} lemma, \cite{Temple}. 
\begin{lemma}\label{div-curl} 
Let $\,M\,$ be a smooth $\,3$-dimensional Riemannian manifold. 
Let $\,\omega_1^m\in L^2\,$ be a sequence of matrix-valued differential forms 
and $\,\omega_2^m\in L^2\,$ be a sequence of matrix-valued differential forms on
$\,M$.  If $\,\omega_1^m\,$ converges weakly in $\,L^2\,$ to a form $\,\omega_1$ 
and $\,\omega_2^m\,$ converges weakly in $\,L^2\,$ to a form $\,\omega_2$, 
and if each sequence $\,d\omega_1^m\,$ and $\,d\omega_2^m\,$ 
is precompact in $\,W^{-1,2}_{loc}(M)$, then 
$\,\omega_1^m\wedge \omega_2^m\,$ converges to $\,\omega_1\wedge \omega_2\,$ 
in the sense of distributions.
\end{lemma} 
In our case, $\,d\,\left(\varphi [a_n,\varphi]\right)\,$ is given by equation (\ref{eq2}). 
We have 
$$
\begin{array}{c}
\|\langle a_n,\,\varphi\rangle\wedge
D_{a_n}\varphi\|_{L^{\frac32}}\,\le\,\|\langle a_n,\,\varphi\rangle\|_{L^6}\;
\|D_{a_n}\varphi\|_{L^2}\,, \\
\| d\varphi\wedge  [a_n,\,\varphi]\,+\, [a_n,\,\varphi]\wedge
d\varphi\|_{L^2}\,\le\,2\|[ a_n,\,\varphi]\|_{L^2}\;
\|d\varphi\|_{L^\infty}\,,
\end{array}
$$
and $\,\|\langle a_n,\,\varphi\rangle\|_{L^6}\,\le\,c\,\|\langle
a_n,\,\varphi\rangle\|_{W^{1,2}}\,$ by the embedding $\,W^{1,2}\subset L^6$. 
We note that $\,L^{\frac32}\,$ and $\,L^{2}\,$ are compactly embedded in $\,W^{-1,2}_{loc}(M)\,$ 
by the Sobolev embedding theorem. Using estimate (\ref{elliptic}), we conclude that the sequence 
$\,d\,\left(\varphi [a_n,\varphi]\right)\,$ is precompact in $\,W^{-1,2}_{loc}(M)$. 
Applying the {\it div - curl} lemma, we obtain that 
$\,\xi\,=\,\varphi\,[a,\varphi]\wedge \varphi\,[a,\varphi]$. 
At this point we can conclude that 
$\,E_\varphi[a]\,\le\,\liminf E_\varphi[a_n]$. It remains to prove that 
$\,a\in {\frak A}_\varphi$. 

We start by showing that the limit form, $\,a$, is distributionally flat. 
We know, that $\,da_n\,$ converges to 
$\,da\,$ in the sense of distributions 
since $\,a_n\rightharpoonup a\,$ in $\,L^2$. 
Next, $\,a_n\wedge a_n\,$ converges to $\,a\wedge a\,$ in distributions since 
$$
a_n\wedge a_n\,=\,\frac14\,[a_n,\varphi]\wedge [a_n,\varphi] 
\,-\,\langle a_n,\,\varphi\rangle\wedge [a_n,\varphi]\,,
$$ 
and the right hand side  converges in the sense of distributions by the previous 
arguments. 

We next note that the holonomy of $\,a\,$ is trivial by Lemma 8 of \cite{AK}. 
We have already seen that the energy of $\,a\,$ is finite. 
The harmonic part of $\,\langle a_n,\,\varphi\rangle\,$ converges to the harmonic 
part of $\,\langle a,\,\varphi\rangle\,$ because the space of harmonic forms 
is finite dimensional. 

Finally, consider the degree,
\[
 \begin{array}{rl}
-\frac{1}{12\pi^2}\int \hbox{Re}\left(a_n\wedge a_n\wedge a_n\right)\,= &\,
-\frac{1}{96\pi^2}\int \hbox{Re}\left(\varphi [a_n,\varphi]\wedge \varphi [a_n,\varphi]
\wedge \varphi [a_n,\varphi]\right)\\ 
 & +\,-\frac{1}{16\pi^2}\int \hbox{Re}\left(\langle a_n,\,\varphi\rangle\varphi\wedge
[a_n,\varphi]\wedge [a_n,\varphi]\right)\,.
\end{array}
\]
We already know that $\,\langle a_n,\,\varphi\rangle\,$ 
converges strongly in $\,L^2\,$ to $\,\langle a,\,\varphi\rangle\,$ 
and $\,\varphi[a_n,\varphi]\wedge \varphi[a_n,\varphi]\rightharpoonup
\varphi[a,\varphi]\wedge \varphi[a,\varphi]$. 
This implies the convergence of the second term. 
We are going to use the {\it div - curl\/} lemma on the first term.
We know that 
$\,\varphi[a_n,\varphi]\rightharpoonup \varphi[a,\varphi]\,$ 
and  $\,d\,\left(\varphi[a_n,\varphi]\right)\,$ 
is precompact in $\,W^{-1,2}$. Now, 
$$
\begin{array}{rl}
d\left([a_n,\varphi]\wedge [a_n,\varphi]\right)\,= &\,
2\, \langle a_n,\,\varphi\rangle\wedge\varphi [D_{a_n}\varphi,\, D_{a_n}\varphi]\,
+\,2\, \langle a_n,\,\varphi\rangle\wedge [\varphi d\varphi,\, D_{a_n}\varphi]\\
 & +\,\varphi d\varphi\wedge [a_n,\varphi]\wedge  [a_n,\varphi]\,.
\end{array}
$$
The first $1$-form factor in each term is uniformly bounded in $\,L^6$ 
and the following $2$-forms are uniformily bounded in $\,L^2$. It follows that the entire
expression is uniformly bounded in $\,L^{\frac32}$, hence precompact in 
$\,W^{-1,2}$. 
The {\it div - curl\/} lemma allows us to conclude that, 
after taking a subsequence, $\,\hbox{cs}(a_n)\to \hbox{cs}(a)$. 
This completes the proof of the theorem.
\hfill$\square$


\section{Minimizers of the Faddeev functional, (II)}
\label{sec:4} 
 
 Given a smooth reference map $\,\varphi$, 
 we now know that there is a minimizer of $\,E_\varphi\,$ in the class $\,\frak A_\varphi$. 
 The smooth connections in this class correspond exactly to $\,\hbox{SO}(3)$-equivalence 
 classes of smooth maps from $\,M\,$ to $\,S^2\,$ homotopic to $\,\varphi$. 
 A general connection, $\,a$,  in the class $\,\frak A_\varphi$ may also be represented as 
 $\,a\,=\,u^*\,du$, but now  $\,u\,$ is only in $\,W^{1,2}(M,\hbox{Sp}_1)$. This follows from 
 Lemma 6 of \cite{AK}. The corresponding map $\,\psi\,=\,u\,\varphi\,u^*\,$ lives 
 in $\,W^{1,2}(M, S^2)\,$ and has finite energy $\,E(\psi)$. 
 We believe that minimizers of $\,E_\varphi\,$ are smooth, but this is an open problem. 
 If a minimizer, $\,a$, is smooth, then the corresponding $\,u\,$ and $\,\psi\,$ are 
 smooth as well.  
 In the smooth case, by Lemma \ref{lemma1}, 
 $\,\psi^*\,\mu_{S^2}\,=\,\varphi^*\,\mu_{S^2}\,$ independent of $\,u$. 
 Even stronger, $\,\psi\,$ is homotopic to $\,\varphi$. 
 Thus, such a $\,\psi\,$ would be a minimizer of the original Faddeev functional, $\,E$, 
 in the class of smooth maps homotopic to $\,\varphi$. 
 Even without any additional regularity result, $\,\psi\,$ is a minimizer of $\,E\,$ 
 in the class of finite energy Sobolev maps of the form $\,u\,\varphi\,u^*\,$ with 
 $\,u\in W^{1,2}(M,\hbox{Sp}_1)$ and 
 $\,\hbox{cs}(u^*\,du)\in \{2(\varphi^*\mu_{S^2}\cup \eta)[M]\,\vert\,\eta\in H^1(M;\mathbb Z)\}$. 
 It is an open problem to extend obstruction theory to
 the class of finite energy Sobolev maps. In particular, 
 a reasonable extension of  the definition of pull-back for Sobolev maps 
 with finite energy would imply $\,\psi^*\,\mu_{S^2}\,=\,\varphi^*\,\mu_{S^2}$. 
 The pull-back by a $\,W^{1,2}(M, S^2)\,$ map is well-defined in the de Rham 
 theory. 
 
 \begin{lemma} For any $\,u\in W^{1,2}(M,\hbox{Sp}_1)$, we have 
 $$
 (u\,\varphi\,u^*)^*\,\omega_{S^2}\,=\,\varphi^*\,\omega_{S^2}\,
 +\,{1\over 2\pi}\,d\,\hbox{Re}\left(\,\varphi\,u^*\,du\right)
 $$
 almost everywhere on $\,M$.
 \end{lemma}
 \noindent{\bf Proof.\/} 
 In quaternionic notation, the standard volume form on $\,S^2\,$ is given by 
 $\,\omega_{S^2}\,=\,-\,\frac{1}{8\pi}\,\hbox{Re}\,(z\,dz\wedge dz)$. 
 The proof follows by straightforward computation.\hfill$\square$  
 
 Notice, that the integrals $\,\int_M \psi^*\,\omega_{S^2}\wedge \eta_k\,$ 
 specify $\,\varphi^*\,\mu_{S^2}\,$ up to torsion. 
 The secondary homotopy invariant works even better. To wit: 
 $\,\int_M\hbox{Re}(u^*\,du\wedge u^*\,du\wedge u^*\,du)\,$ is
 well defined. In addition to Theorem \ref{theorem3}, we prove the following result. 
 
 Let $\,\frak F_k\,$ denote the class of all finite energy 
 maps $\,\psi\in W^{1,2}(M, S^2)\,$  for which there exist a  $1$-form 
 $\,\theta^\psi \in W^{1,2}\,$ 
 $$
 \psi^*\,\omega_{S^2}\,=\,d\,\theta^\psi
 $$
 and 
 \begin{equation}\label{hopf2}
 \int_M \theta^\psi\,\wedge d\,\theta^\psi\,=\,k\,.
 \end{equation} 
 The argument from \cite{VK} shows, that there exists a constant $\,c_M>0\,$ 
 so that 
 $$
 E(\psi)\,\ge\,c_M\,|k|^{\frac34}\,,
 $$
 for all $\,\psi\in\frak F_k$.
 
 \begin{theorem} \label{theorem4}
 In every nonvoid class $\,\frak F_k$ there exists a minimizer of $\,E$.
 \end{theorem}
 \noindent{\bf Proof.\/} Choose a minimizing sequence, $\,\psi_n$, and let 
 $\,\theta_n\,$ denote the corresponding $\,\theta^{\psi_n}$. By Hodge theory, we 
 may assume that each $\,\theta_n\,$ is co-closed ($\,\delta\,\theta_n\,=\,0$), 
 and has trivial harmonic part. 
 Taking a subsequence,  
$\,\psi_n\,$ converges weakly in $\,W^{1,2}(M, S^2)\,$ and almost everywhere 
to some $\,\psi\in W^{1,2}(M, S^2)$. 
A direct computation shows that pointwise
$\,|\psi^*\,\omega_{S^2}|\,=\,(8\pi)^{-1} \,|d\psi\wedge d\psi|$. Since $\,E(\psi_n)\,$ is bounded, 
$\,\|\psi_n^*\,\omega_{S^2}\|_{L^2}\,$ is bounded. 
Taking another subsequence, 
$\,\psi_n^*\,\omega_{S^2}\,\rightharpoonup\,\xi\,$ in $\,L^2$, for some $\,\xi$. To show that 
$\,\xi\,=\,\psi^*\,\omega_{S^2}$, we use the {\it div - curl} lemma. We have 
$\,\psi_n^*\,\omega_{S^2}\,=\,-\frac{1}{8\pi}\,\hbox{Re}(\psi_n\, d\psi_n\wedge d\psi_n)$. 
Certainly, $\,d\psi_n\rightharpoonup d\psi\,$ and 
$\,\psi_n\,d\psi_n\rightharpoonup \psi\,d\psi\,$ in $\,L^2$. Their differentials are 
$0$ and $\,d\psi_n\wedge d\psi_n$,  respectively, which are both precompact in 
$\,W^{-1,2}$. Thus, $\,\xi\,=\,\psi^*\,\omega_{S^2}$. This implies that 
$\,E(\psi)\,\le\,\inf\limits_{\frak F_k} E$. 
The $1$-forms $\,\theta_n\,$ are  uniformly bounded in 
$\,W^{1,2}$, hence, by taking a subsequence, converge weakly in $\,W^{1,2}\,$  
and strongly in $\,L^2\,$ to 
some $\,\theta$. At the same time, $\,d\theta_n\rightharpoonup d\theta\,$ in $\,L^2$. 
Recalling that $\,d\theta_n\,=\,\psi_n^*\,\omega_{S^2}\rightharpoonup \psi^*\,\omega_{S^2}$, 
we conclude, that $\,d\theta\,=\, \psi^*\,\omega_{S^2}$. Since $\,\theta_n\to \theta\,$ 
and $\,d\theta_n\rightharpoonup d\theta$, we obtain $\,\int_M \theta\wedge d\theta\,=\,k$. 
Thus, $\,\psi\in \frak F_k\,$ and $\,\psi\,$ is the minimizer of $\,E$. \hfill$\square$


\section{Concluding remarks}
\label{sec:5} 

One can pose many minimization problems for the  functionals
$$
E(\psi)\,=\,\int_M |d\psi|^2\,+\,|d\psi\wedge d\psi|^2\,d\hbox{vol}\,,
$$
and 
$$
E_{\varphi}[a]\,:=\,\int_M |D_a \varphi|^2\,+\,
|D_a \varphi\wedge D_a \varphi|^2\,.
$$ 
For example, one could minimize the first functional over all maps with fixed 
primary obstruction only; one could minimize the second functional over 
the classes of flat connections with fixed nontrivial holonomy, or fixed Chern-Simons 
invariant, or arbitrary holonomy. 
The arguments that we used in Theorems 3 and 4 
apply equally well to each of these problems. 

There are several interesting open questions related to this model. 
We have already mentioned, that we expect the minimizers to be smooth. 
What about maps in general: 
Do  $\,W^{1,2}\,$ maps with finite Faddeev energy have extra regularity? 
Note that the second term in $\,E(\psi)\,$ rules out local singularities 
of the form $\,x\mapsto {x\over |x|}$. 
How does one extend obstruction theory to finite energy maps? 
In particular, what is the appropriate definition of homotopy of
finite energy maps? Is there a cohomology theory that agrees with 
integral singular theory  for which pull-back is defined for 
finite energy maps? Pull-backs in such a theory should also be 
homotopy invariant and analogues of Lemma \ref{lemma1} and 
Proposition 1 should hold. (Note that the proofs of  
Lemma \ref{lemma1} and 
Proposition 1 only require continuity. Thus, the above questions only 
make sense if there exist honestly discontinuous finite energy maps.) 
The Hopf invariant given in equation (\ref{hopf}) or $\,k\,$ in (\ref{hopf2}) 
should be an integer. This has not been verified for Sobolev maps. 
In another direction, it would be very interesting to see the  structure 
of the minimizers. There may very well exist explicit minimizers on special 
Riemannian manifolds such as the three-sphere, three-torus, lens spaces, etc. 
There may be new phenomena for closed domains that could be discovered 
by numerical experimentation. The closed case has the additional advantage that
one does not have to worry about behavior at infinity.

\end{document}